\documentclass[twocolumn,showpac,aps,prb,superscriptaddress]{revtex4-1}
\usepackage{color}
\usepackage{epsfig}
\usepackage{graphicx}
\usepackage{dcolumn}
\usepackage{bm}
\usepackage{amssymb}
\usepackage{amsmath}
\usepackage{tipa}
\begin{document}

\title{Bulk electronic structure of non-centrosymmetric Eu$T$Ge$_3$ ($T$= Co, Ni, Rh, Ir) studied by hard x-ray photoelectron spectroscopy}

\author{Yuki Utsumi}
\email{Yuki.Utsumi@synchrotron-soleil.fr}
\affiliation{Synchrotron SOLEIL, L'Orme des Merisiers, BP 48, St Aubin, 91192 Gif sur Yvette, France}

\author{Deepa Kasinathan}
\affiliation{Max Planck Institute for Chemical Physics of Solids, N\"othnitzer Stra\ss e 40, 01187 Dresden, Germany}

\author{Przemys\l aw Swatek}
\affiliation{Division of Materials Science and Engineering, Ames Laboratory, Ames, 50011 Iowa, USA}
\affiliation{Department of Physics and Astronomy, Iowa State University, Ames, 50011 Iowa, USA}
\affiliation{Institute of Low Temperature and Structure Research, Polish Academy of Sciences, P.O. Box 1410, 50-950 Wroclaw, Poland}

\author{Oleksandr Bednarchuk}
\affiliation{Institute of Low Temperature and Structure Research, Polish Academy of Sciences, P.O. Box 1410, 50-950 Wroclaw, Poland}

\author{Dariusz Kaczorowski}
\affiliation{Institute of Low Temperature and Structure Research, Polish Academy of Sciences, P.O. Box 1410, 50-950 Wroclaw, Poland}

\author{James M. Ablett}
\affiliation{Synchrotron SOLEIL, L'Orme des Merisiers, BP 48, St Aubin, 91192 Gif sur Yvette, France}

\author{Jean-Pascal Rueff}
\affiliation{Synchrotron SOLEIL, L'Orme des Merisiers, BP 48, St Aubin, 91192 Gif sur Yvette, France}
\affiliation{Laboratoire de Chimie Physique-Mati\`{e}re et Rayonnement, Universit\'{e} Pierre et Marie Curie-Paris 6, CNRS-UMR7614, 11 Rue Pierre et Marie Curie, 75005 Paris, France}

\date{\today}

\begin{abstract}
Non-centrosymmetric Eu$T$Ge$_3$ ($T$=Co, Ni, Rh, and Ir) possesses magnetic Eu$^{2+}$ ions and antiferromagnetic ordering appears at low temperatures. Transition metal substitution leads to changes of the unit cell volume and of the magnetic ordering. However, the magnetic ordering temperature does not scale with the volume change and the Eu valence is expected to remain divalent. 
Here we study the bulk electronic structure of non-centrosymmetric Eu$T$Ge$_3$ ($T$=Co, Ni, Rh, and Ir) by hard x-ray photoelectron spectroscopy. The Eu 3$d$ core level spectrum confirms the robust Eu$^{2+}$ valence state against the transition metal substitution with a small contribution from Eu$^{3+}$. The estimated Eu mean-valence is around 2.1 in these compounds as confirmed by multiplet calculations. In contrast, the Ge $2p$ spectrum shifts to higher binding energy upon changing the transition metal from 3$d$ to 4$d$ to 5$d$ elements, hinting of a change in the Ge-$T$ bonding strength. The valence bands of the different compounds are found to be well reproduced by {\it ab initio} band structure calculations. 
\end{abstract}
\maketitle

\section{Introduction}
Strongly correlated 4$f$-electron systems have been a platform for studying various anomalous phenomena, such as valence fluctuations, unconventional superconductivity, heavy fermion behavior and spin/charge ordering \cite{Varma1976, Antonov2011}.
The ground state property of these compounds are characterized by competing Kondo effects or Rudermann-Kittel-Kasuya-Yoshida (RKKY) interactions. Both interactions originate from the interplay of localized $f$ electrons and itinerant conduction electrons, though the former quenches the magnetic moments, while the latter leads to magnetic ordering in the ground state. The competition between Kondo effect and RKKY interactions in Ce- and Yb-compounds are often discussed within the Doniach phase diagram \cite{Doniach1977}. In the vicinity of the quantum critical point (QCP), where the non-thermal parameter controlled phase transition happens at absolute zero temperature, particularly, quantum fluctuations accommodate exotic phenomena \cite{Si2010}.

Eu-compounds exhibit very different phase diagrams from Ce- and Yb-compounds, and an absence of a QCP. Most of the reported Eu-compounds favor a Eu$^{2+}$ ($4f^7$, $J$=7/2) valence state with an antiferromagnetic ground state. However, the energy difference between Eu$^{2+}$ and the non-magnetic Eu$^{3+}$ ($4f^6$, $J$=0) valence state is not so large\cite{Bauminger1973} and is reachable by applying external pressure or chemical substitution.
Indeed, amongst the most extensively studied Eu-compounds series with the ThCr$_2$Si$_2$-type crystal structure, pressure or chemical substitution controlled first-order valence transitions and valence fluctuations are frequently reported \cite{Onuki2016}. In the Eu(Pd$_{1-x}$Au$_x$)$_2$Si$_2$ system, EuAu$_2$Si$_2$ possesses a Eu$^{2+}$ valence state and exhibits antiferromagnetic ordering below the N\'eel temperature ($T_{\rm N}$) of $\sim15.5$ K \cite{Felner1975}. Substitution of smaller Pd ions decreases the lattice parameter and, by contrast, increases $T_{\rm N}$. Above $x\sim0.25$, the magnetic transition is suddenly taken over by a first-order valence transition to Eu$^{3+}$.\cite{Segre1982} The Eu valence deviates from integer values, to so called intermediate valence states, and is $\sim$2.8 in EuPd$_2$Si$_2$ below 150 K \cite{Sampath1981}. A similar tendency is also reported for Eu(Pt$_{1-x}$Ni$_x$)$_2$Si$_2$ \cite{Mitsuda2007} and EuNi$_2$(Si$_{1-x}$Ge$_x$)$_2$ \cite{Wada1999} systems in such a way that the substitution by elements with small ionic radii works in the same way as pressure and leads to a non-magnetic ground state. Application of external pressure shows a consistent behavior compared with chemical substitutions \cite{Hesse1997}. Due to the different ionic size between Eu$^{2+}$ and Eu$^{3+}$, the change of the Eu valence state is often assigned to Kondo volume collapse effects\cite{Allen1982, Allen1992}.
In contrast, changes of the Eu valence state and ground state property are found to be independent of the volume effect in Eu(Rh$_{1-x}$Ir$_x$)$_2$Si$_2$ system. The conversion from divalent EuRh$_2$Si$_2$ to valence-fluctuating EuIr$_2$Si$_2$ involves only a $\sim$1.5\% volume change, indicating its origins in electronic structure changes\cite{Seiro2011}.
Very recently, exotic behavior has been discovered in EuRhSi$_3$ and Eu$_2$Ni$_3$Ge$_5$ \cite{Nakamura2015, Muth2016, Nakashima2017} that cannot be explained by the conventional phase diagram of Eu-compounds. Both EuRhSi$_3$ and Eu$_2$Ni$_3$Ge$_5$ have magnetic Eu$^{2+}$ ions and exhibit antiferromagnetism below $T_{\rm N}$=49\,K and 19\,K, respectively, at ambient pressure. Electrical resistivity measurements under pressure have reported the suppression of a magnetic ordering temperature and a successive phase transition to a non-magnetic heavy fermion state without a hint of a valence transition \cite{Muth2016, Nakashima2017}. These behaviors are similar to Ce- and Yb-compounds and indicate the possible existence of a QCP. 
These new aspects of Eu-compounds urge a systematic study of the electronic structure and its relation to physical properties.

Eu-based ternary germanides Eu$T$Ge$_3$ ($T$=Co, Ni, Rh, Ir) and EuRhSi$_3$ are isostructural, possess a BaNiSn$_3$-type structure ($I4mm$) which is similar to the ThCr$_2$Si$_2$-type structure, though without centrosymmetry \cite{Bednarchuk2015a}. Magnetic susceptibility measurements \cite{Goetsch2013, Bednarchuk2015a, Bednarchuk2015b, Bednarchuk2015c} and M\"ossbauer spectroscopy \cite{Maurya2014, Maurya2016} report the presence of magnetic Eu$^{2+}$ ions in all the compounds and localized Eu 4$f$ moments order antiferromagnetically at similar temperatures. The magnetic moments order antiferromagnetically along the $c$-axis at $T_{\rm N}$=15.4, 13.5 and 12.3 K for EuCoGe$_3$, EuNiGe$_3$ and EuIrGe$_3$, respectively. EuCoGe$_3$ and EuIrGe$_3$ exhibit additional magnetic transitions at 13.4 and 7.5 K, respectively, due to a change of the antiferromagnetic structure\cite{Bednarchuk2015a, Bednarchuk2015c}. Recently, three antiferromagnetic phases have been discovered in EuIrGe$_3$ and a helical magnetic structure based on the Dzyaloshinskii-Moriya interaction was proposed.\cite{Kakihana2017} Conversely in EuRhGe$_3$, the magnetic moments order perpendicular to the $c$-axis at $\sim$12 K \cite{Bednarchuk2015a, Bednarchuk2015c}. 
For each compound, the effective magnetic moments are close to the Eu$^{2+}$ ionic value of 7.90 $\mu_{\rm B}$ \cite{Bednarchuk2015a}. All the transition metals are nonmagnetic in Eu$T$Ge$_3$. Despite the variation of transition metal substitution and the change in the unit cell volume, the Eu ions seems to have a robust Eu$^{2+}$ valence state with $T_{\rm N}$ being barely affected. 
In the EuNi(Si$_{1-x}$Ge$_x$)$_3$ system, transport measurements reported a monotonous decrease of $T_{\rm N}$ with an increase in Ge substitution indicating its strong connection with the volume change \cite{Uchima2014a}.
However, the change of $T_{\rm N}$ in Eu$T$Ge$_3$ by transition metal substitution does not show a propotional change with the unit cell volume. This implies that variation of the physical properties of the Eu$T$Ge$_3$ by transition metal substitution is rather dominated by the change in electronic structure rather than the unit cell volume effect.
In order to study the transition metal substitution effect on the electronic structure of Eu$T$Ge$_3$, we performed hard x-ray photoelectron spectroscopy (HAXPES). By using the bulk sensitive HAXPES method, we can unambiguously determine the Eu valence from Eu 3$d$ core level spectra and suppress the surface contribution. The Eu 3$d$ core level spectra confirmed that the Eu$^{2+}$ valence state is robust against transition metal substitution. The estimated Eu valence is close to 2.1. In contrast, Ge 2$p$ core level spectrum shifts to high binding energy by changing transition metal from 3$d$ to 4$d$ to $5d$ elements. A similar trend was observed in the Eu 4$f$ spectrum in the valence band. We compare the measured valence band electronic structure with {\it ab initio} band structure calculations.

\begin{figure}[t!]
\includegraphics[width=0.5\textwidth]{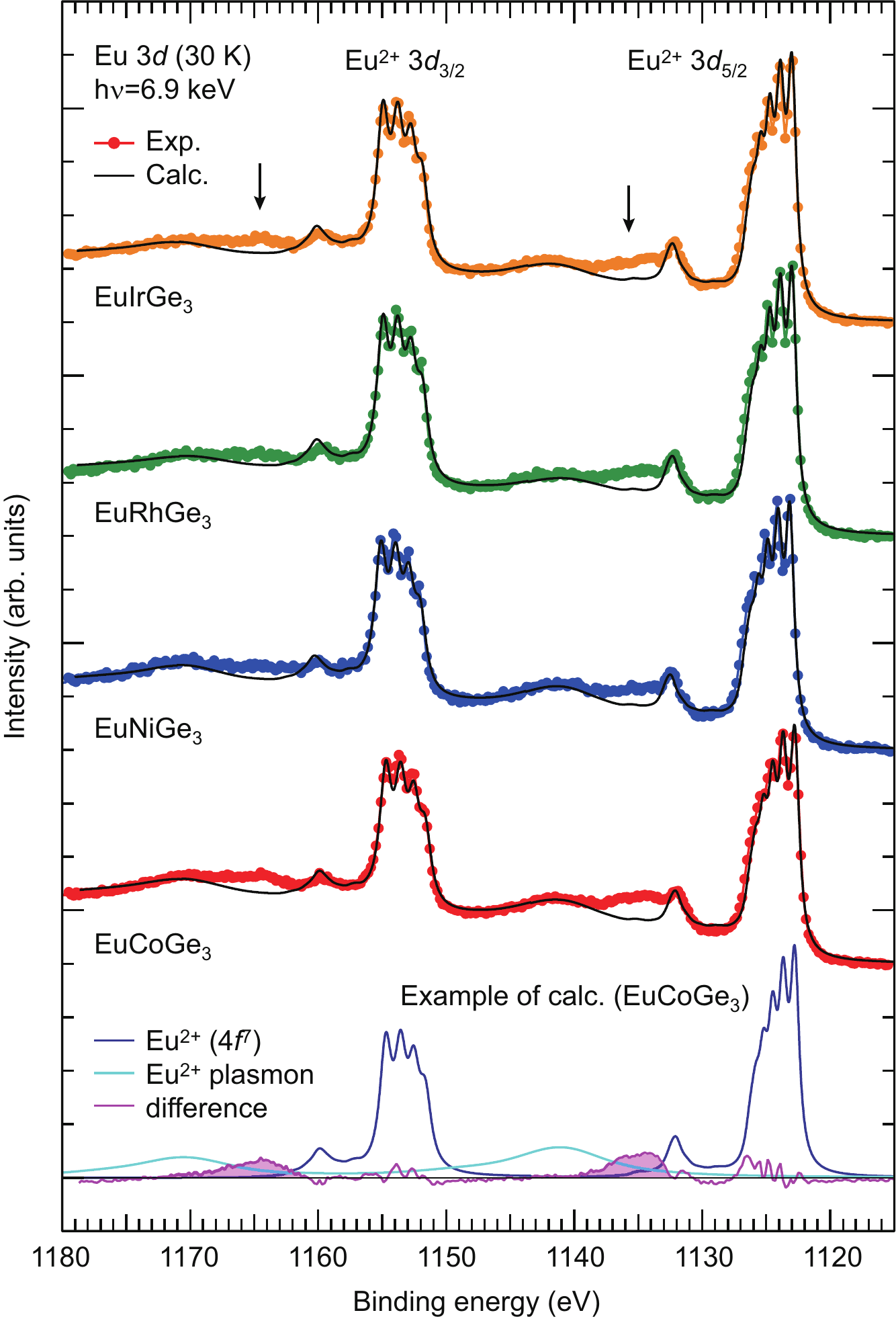}
\caption{(Color online).
Eu 3$d$ core level spectra of Eu$T$Ge$_3$ ($T$=Co, Ni, Rh and Ir). The experimental data are displayed using filled circles. The solid black lines represent the simulated spectra from atomic multiplet calculations with a 4$f^7$ configuration including plasmon satellites and an integral background \cite{Shirley1972}. The arrows indicate the position of Eu$^{3+}$ components. The simulated Eu$^{2+}$ spectrum (dark blue line) and its plasmon satellites (light blue line) for EuCoGe$_3$ are shown in the bottom of the figure as an example. The difference (purple line) spectrum obtained by subtracting the simulated spectrum from the experimental one reveals the Eu$^{3+}$ components (purple area).
}
\label{fig1}
\end{figure}

\section{Experimental}
HAXPES measurements were performed at the GALAXIES beamline \cite{Ceolin2013, Rueff2015} of the SOLEIL synchrotron. The incident energy was selected by using the third order of the Si(111) monochromator ($h\nu$=6.9 keV) yielding a photon bandwidth of $\sim 200$ meV. The photon beam was linearly polarized with the electrical field vector in the plane of the storage ring. Photoelectrons were collected by using a hemispherical analyzer EW4000 (VG Scienta). The binding energy of spectra was calibrated by measuring the Fermi edge of a Au film. The overall energy resolution was estimated to be  $\sim$250 meV from Au Fermi edge fitting.
Eu$T$Ge$_3$ ($T$=Co, Ni, Rh and Ir) single crystals were grown by the metal-flux method \cite{Bednarchuk2015a}. The grown crystals were characterized by x-ray diffraction, magnetic susceptibility and electrical resistivity measurements.
The clean surfaces of the samples were obtained by fracturing {\it in-situ} under vacuum (better than 5$\times$10$^{-8}$ mbar). The samples were aligned in a grazing incidence (normal emission) geometry. In order to avoid radiation damage, all the measurements were performed at the lowest reachable temperature of 30 K.

\begin{figure*}[t!]
\includegraphics[width=\textwidth]{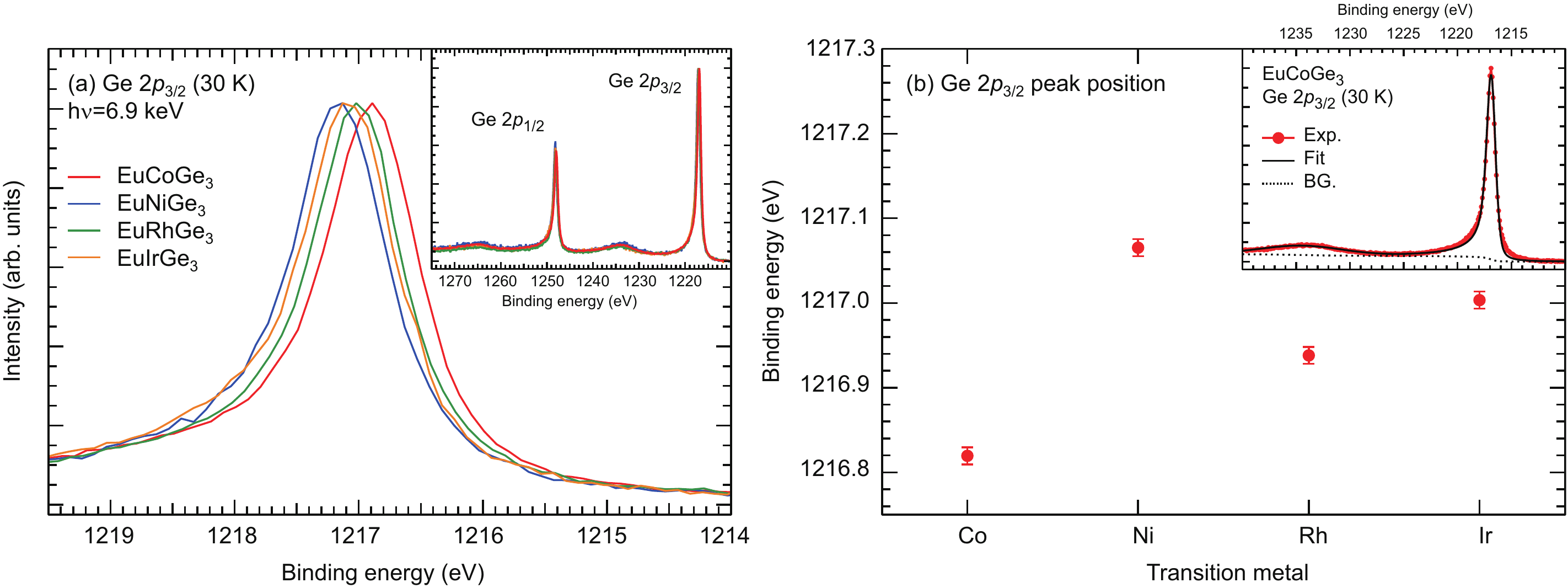}
\caption{(Color online) (a) Ge 2$p_{3/2}$ core level spectra measured at 30 K. The inset shows the complete Ge 2$p$ spectra. (b) Binding energy of the Ge 2$p_{3/2}$ peaks of Eu$T$Ge$_3$ as a function of $T$ ($T$=Co, Ni, Rh, and Ir). The inset shows an example of the fit. The dashed line represents the integral background \cite{Shirley1972} (BG).}

\label{fig2}
\end{figure*}

\section{Results and discussion}
Figure \ref{fig1} shows the Eu 3$d$ core level spectra of Eu$T$Ge$_3$ ($T$=Co, Ni, Rh and Ir) measured at 30 K. The Eu 3$d$ spectra are split into a 3$d_{5/2}$ (1120-1145 eV) and 3$d_{3/2}$ (1150-1175 eV) components due to spin-orbit interaction. Each spin-orbit partner further split into a Eu$^{2+}$ component at lower binding energy and the Eu$^{3+}$ component at higher binding energy representing the Eu $4f^7 \rightarrow \underline{c}~4f^7 + e$ and Eu $4f^6 \rightarrow \underline{c}~4f^6 + e$ transitions, respectively. Here, $\underline{c}$ denotes a $3d$ core hole and $e$ the outgoing photoelectron. Broad structures around 1140 and 1170 eV are attributed to plasmon statellites related to the Eu$^{2+}$ 3$d$ photoemission process. Compared to the Eu 3$d$ spectrum of other divalent Eu compounds, such as EuRh$_2$Si$_2$ (mean-Eu valence $v\sim2.1$ at 300-20 K) \cite{Ichiki2017} and EuNi$_2$(Si$_{0.21}$Ge$_{0.79}$)$_2$ ($v\sim2.2$ at 300 K) \cite{Ichiki2017b}, the relative intensity of Eu$^{3+}$ components to those of Eu$^{2+}$ is small and buried in the tail of the Eu$^{2+}$ components and its statellite structures. As expected from magnetic susceptibility and M\"ossbauer measurements, the Eu valence states in the Eu$T$Ge$_3$ series are very close to Eu$^{2+}$. In order to elucidate Eu$^{3+}$ contributions, a simulation analysis was performed by carrying out atomic multiplet calculations to account for the lineshape of the Eu 3$d$ core level spectra. The Eu 3$d$ spectra were simulated by using the XTLS (version 9.01) code \cite{Tanaka1994} with a 4$f^{7}$ (Eu$^{2+}$) ground-state configuration. The electrostatic and exchange parameters were obtained by Cowan's atomic Hartree-Fock program with relativistic corrections \cite{Cowan1981}. The exchange parameters were scaled down to 86$\%$ of their Hartree-Fock values. The calculated spectra are convoluted with a Lorentzian function for lifetime broadening and a Gaussian to account for the experimental resolution. The broadening parameters as well as the values used for the Coulomb and exchange multiplet interactions are listed in Ref. \onlinecite{Slater}. An example of the simulation for EuCoGe$_3$ is shown in the bottom of Fig. \ref{fig1}. The plasmon satellites (light blue line) are reproduced by broadening the simulated Eu$^{2+}$ atomic multiplet spectrum and shifting in order to be in agreement with the experimental energy. Their relative intensity and the energy position to the Eu$^{2+}$ 3$d$ components were calibrated using the Ge 2$p$ peak and its plasmon position (see inset of Fig. \ref{fig2} (b)). The solid black lines in Fig. \ref{fig1} represent the simulated spectra including the atomic multiplet spectrum, plasmon satellites and integral background \cite{Shirley1972}. The experimental spectra are fitted by adjusting the intensity of the calculated spectra such that the difference between the experimental and the calculated spectra are minimized. As seen in Fig. \ref{fig1}, the simulations can well reproduce the multiplet structures of the experimental spectra.  Since the simulations only take into account the Eu$^{2+}$ contribution, the deviations from the simulated spectrum at $\sim$1135 and 1165 eV are assigned to the Eu$^{3+}$ contributions. We extracted the Eu$^{3+}$ component by subtracting the simulated spectrum from that of experiment. A contribution from the Eu$^{3+}$ plasmon satellites to the Eu $3d$ spectrum is negligibly small and therefore not considered in this analysis. Some residual wiggling feature on the difference spectrum (purple line) originates mostly from tiny deviations in the peak positions and peak widths of the multiplet structures. The Eu valence was estimated by using the formula $v=2+I_{3+}/(I_{2+}+I_{3+}$). Here, $I_{2+}$ and $I_{3+}$ denote integrated spectral intensities of the simulated Eu$^{2+}$ spectrum (dark blue line) and the extracted Eu$^{3+}$ component (purple area in Fig. \ref{fig1}), respectively. The obtained Eu valences are $v$=2.11, 2.09, 2.08 and 2.09 ($\pm0.01$) for EuCoGe$_3$, EuNiGe$_3$, EuRhGe$_3$ and EuIrGe$_3$, respectively. We should note that the estimated inelastic mean-free path is $\sim73$ \AA \ for 5.7 keV photoelectrons \cite{Tanuma1994}, therefore, the Eu$^{3+}$ signal is not likely to be coming from the surface states.

\begin{figure*}[h!t!]
\includegraphics[width=\textwidth]{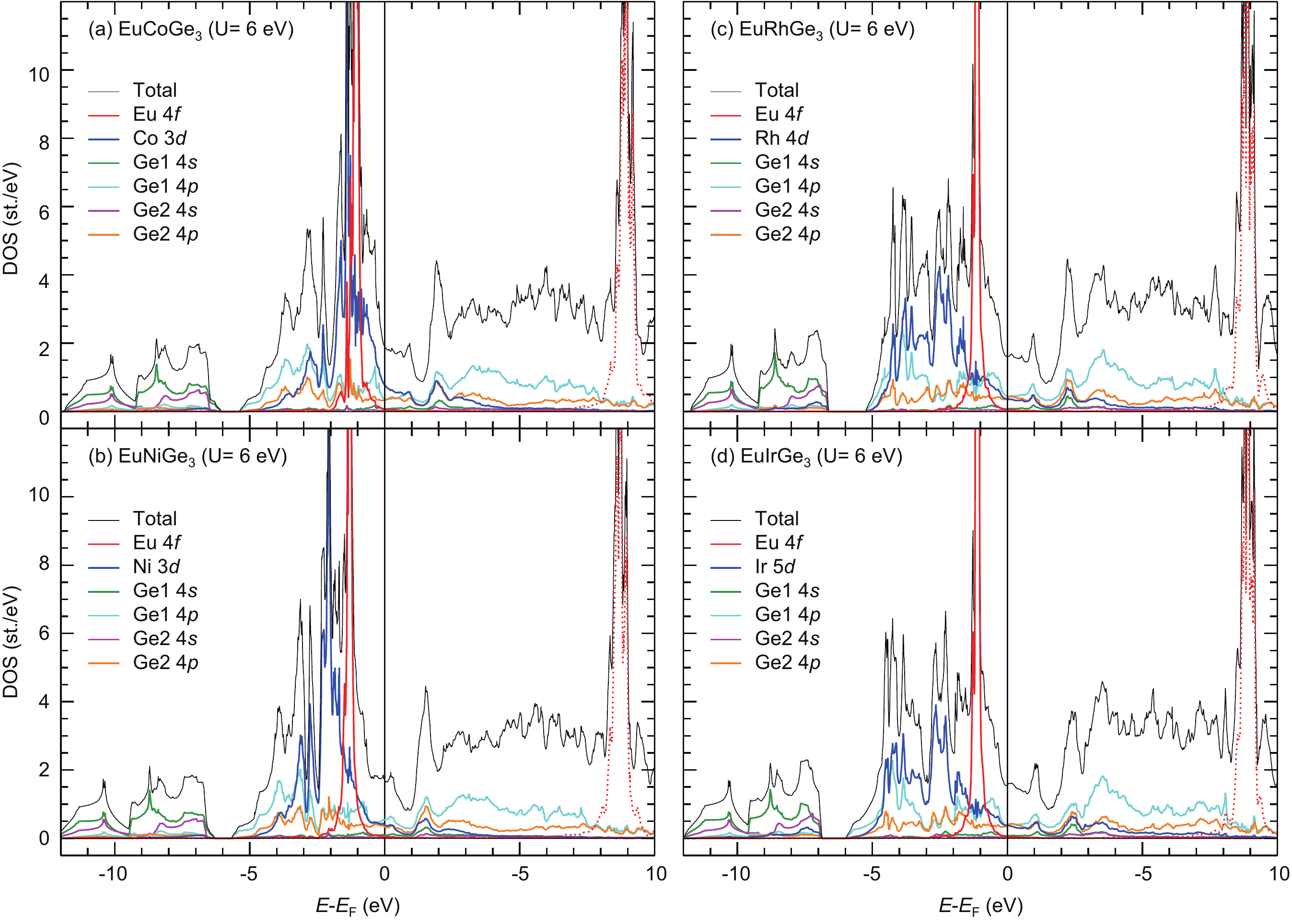}
\caption{(Color online) Total and partial density of states (DOS) of (a) EuCoGe$_3$, (b) EuNiGe$_3$, (c) EuRhGe$_3$ and (d) EuIrGe$_3$ simulated by LDA+$U$ ($U$=6 eV) method. Since all the compounds show symmetric DOS for two spin channels, the DOSs in minority-spin channel are inverted and added together with the majority-spin DOSs, except Eu 4$f$ DOS. The majority-spin DOS of Eu 4$f$ (solid red line) are fully occupied and the minority-spin DOS (dashed red line) are unoccupied.}
\label{fig3}
\end{figure*}

Figure \ref{fig2}(a) shows spectra with the Ge $2p_{3/2}$ component for all the compounds. The inset shows the Ge 2$p$ complete spectra. The Ge 2$p$ spectra tends to shift towards higher binding energy  upon going from 3$d$ to 4$d$  to 5$d$ elements.
In order to obtain a precisely this energy shift, a fitting analysis was performed on the spectra by using Gaussian and Lorentzian functions \cite{note1}. In addition, a Mahan function \cite{Mahan1975} with $\alpha$\,=\,0.16 is used to account for the asymmetry of the line shapes. The result of the fitting analysis is plotted in Fig. \ref{fig2} (b) with an example of such a fit in the inset. A large energy shift of $\sim250$ meV is observed between EuCoGe$_3$ and EuNiGe$_3$. The energy shift of the Ge 2$p$ peak for EuRhGe$_3$ and EuIrGe$_3$ relative to EuCoGe$_3$ are $\sim120$ and 180 meV, respectively. We note that the Fermi level ($E_{\rm F}$) was carefully checked just before and after each measurement to correct for possible drift of the incident photon energy. The energy shift between EuCoGe$_3$ and EuNiGe$_3$ can be understood as a result of the energy shift of $E_{\rm F}$ towards higher-energy in the conduction bands with increasing 3$d$ occupation. 

In the related ternary compounds in the form of $RT_2$Ge$_2$ ($R$: Rare earth, $T$: transition metal), the strong $T$-Ge bonding based upon hybridization of $T-d$ states with Ge$-sp$ states is reported \cite{Jeon1989, Chen1993}. 
In a similar vein, a strong bonding between $T$-Ge can also be expected in the Eu$T$Ge$_3$ family as well. The energy shift of the Ge $2p$ binding energy amongst EuCoGe$_3$, EuRhGe$_3$ and EuIrGe$_3$ indicates the change of bonding by transition metal substitution.

In order to study the atomic orbital character of the valence band electronic structure, we performed band structure calculations using the full-potential nonorthogonal local orbital code (FPLO) \cite{Koepernik1997,Koepernik1999}. The local density approximation (LDA) with the Perdew and Wang flavor \cite{Perdew1992} of the exchange and correlation potential was chosen. Additionally, the strong Coulomb replusion between the Eu 4$f$ electrons of europium was included in a mean-field way by applying the LDA+U method. The calculations were performed for the experimentally obtained lattice parameters reported in Ref. \onlinecite{Bednarchuk2015a} with $J_{\rm H}$=0.7 eV and varying $U$ from 5 to 7 eV. It should be noted that varying $U$ only changes the energy separation between the filled and unfilled Eu 4$f$ states and does not change the results qualitatively. The calculated total and partial density of states (PDOS) are presented in Fig. \ref{fig3}. 
The europium ions in the unit cell are configured such that they have a ferromagnetic arrangement in the $ab$-plane and antiferromagnetically aligned along the $c$-axis. Therefore,  
the DOSs are symmetric for both the spin channels except for Eu 4$f$.
 The DOSs in the minority-spin channel are inverted and added together with the majority-spin DOSs for comparison to experiments. The majority-spin states of Eu $4f$ (solid red line) are fully occupied and appear as a localized sharp peak around 1 eV while the Eu 4$f$ minority-spin states (dashed red line) remain unoccupied. These results reflect the magnetic Eu 4$f^7$ state in all compounds. The Co and Ni 3$d$ PDOSs appear centered at -1.5 eV and -2 eV, respectively. An increase of the 3$d$ electron number in EuNiGe$_3$ shifts the center of 3$d$ PDOS away from $E_{\rm F}$ that decreases the hybridization to Eu $4f$. The Rh 4$d$ and Ir $5d$ PODSs show a more extended nature than 3$d$ PDOS, appear from $E_{\rm F}$ to -6 eV and then below -7 eV. The occupied Ge $4p$ PDOS are mainly distributed from $E_{\rm F}$ to -6 eV, hybridizing with transition metal $d$ and Eu $4f$ PDOSs. The Ge 4$s$ PDOS appears at -7 to -10 eV. Commonly in all four compounds, a quasi-gap-like low DOS region appears $\sim$1 eV above $E_{\rm F}$.

\begin{figure}[h!t!]
\includegraphics[width=0.5\textwidth]{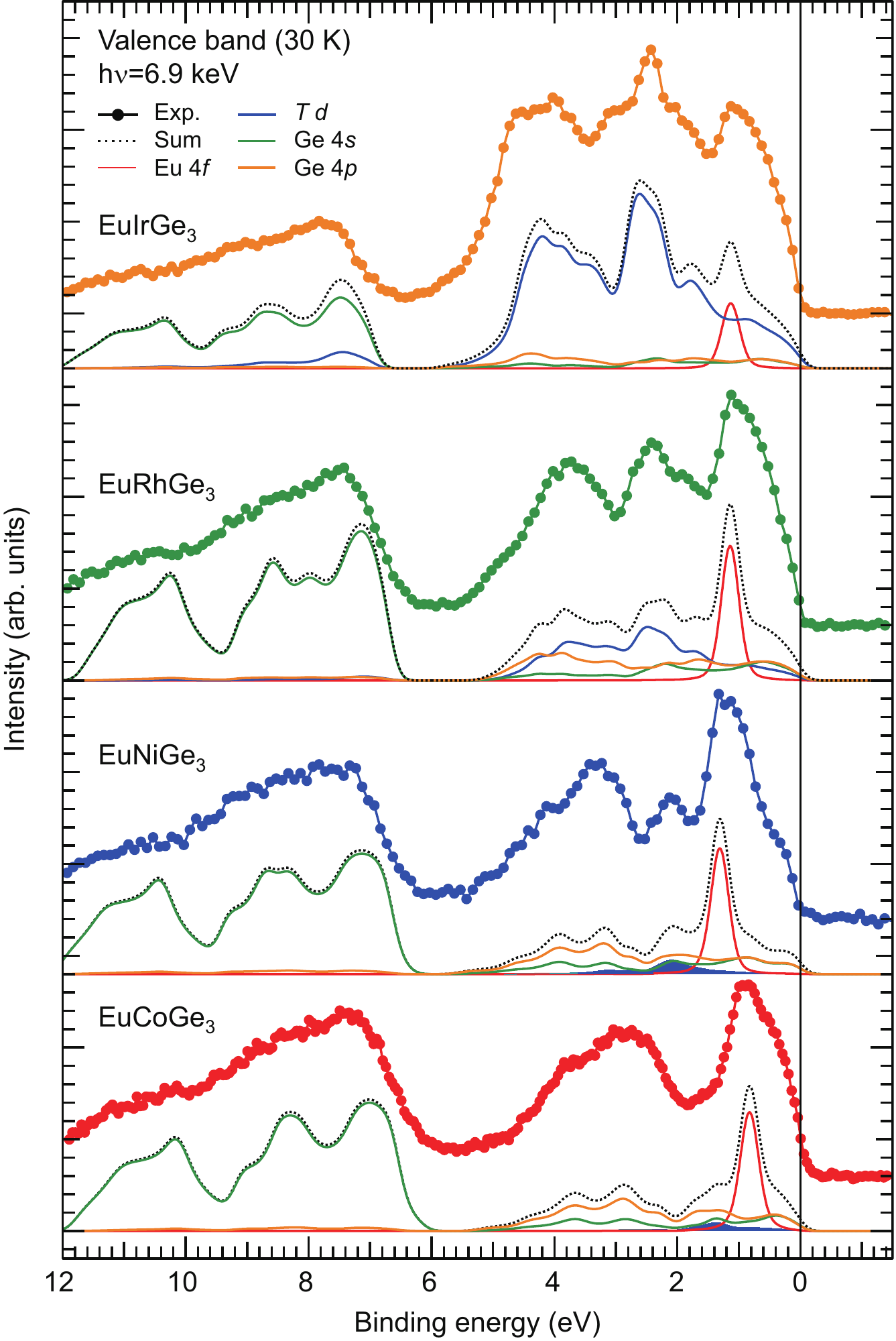}
\caption{(Color online) Valence band spectra (filled circles) of Eu$T$Ge$_3$ ($T$=Co, Ni, Rh and Ir) after integral background correction \cite{Shirley1972} and simulated spectra (lines). The majority- and minority-spin DOSs are added together to compare with experimental spectra. Each PDOSs are multiplied by the corresponding photoionization cross-section and convoluted by a Gaussian (0.3 eV: FWHM) and Fermi-Dirac functions at 30 K. The dashed lines denote the sum of the displayed PDOSs. Note that the intensities of simulated EuIrGe$_3$ spectra are reduced to 50\% relative to those of other compounds.}
\label{fig4}
\end{figure}
The experimentally measured valence band spectra after integral background correction \cite{Shirley1972} are displayed in Fig. \ref{fig4}. The HAXPES valence band spectra cannot be directly compared to the theoretical DOSs, since the photoionization cross-section of Eu 4$f$ states is not the only one to contribute to the spectrum. Therefore the PDOSs are weighted with the corresponding photoionization cross-sections extracted or interpolated from Ref. \onlinecite{Trzhaskovskaya2001, Trzhaskovskaya2002, Trzhaskovskaya2006}. Then, the simulated theoretical spectra were convoluted with a Gaussian function (FWHM: 0.3 eV) and Fermi-Dirac function of 30 K. We consider the Eu 4$f$, $T$ $d$, and Ge $4sp$ states as mainly contributing to the valence band. The simulated spectra show good accordance with experimental spectra. Note that we used the DOS with $U$=5 eV for EuCoGe$_3$ and $U$=6 eV for other compounds, based on the agreement of Eu 4$f$ peak position. In all the compounds, a localized Eu$^{2+}$ 4$f$ peak is observed around 1 eV. A broad round shape of the Eu 4$f$ peak is due to Eu$^{2+}$ multiplet structures \cite{Gerken1983}. A peak around 2 eV in EuNiGe$_3$, EuRhGe$_3$ and EuIrGe$_3$ corresponds to transition metal $d$ states overlapping with the Ge $4sp$ states.
Due to the larger photoionization cross-section of Ir 5$d$ and Rh 4$d$ compared to those of Ni 3$d$ and Co 3$d$, the spectral weight of transition metal $d$ states is enhanced in EuIrGe$_3$ and EuRhGe$_3$.
The broad structure between 6-12 eV is mainly attributed to Ge $4s$ states. The Eu $4f$ peak shows a similar tendency as that of the Ge 2$p$ spectrum, shifting to higher binding energy by changing transition metal atoms. However, unlike the Ge 2$p$ spectrum, no significant energy shift was observed between EuRhGe$_3$ and EuIrGe$_3$.
The experimental valence band spectra have low DOS at $E_{\rm F}$, especially for EuNiGe$_3$. The result is consistent with the reported transport measurements \cite{Uchima2014b}. The decrease in DOS at $E_{\rm F}$ from EuCoGe$_3$ to EuNiGe$_3$ can be understood as a rigid-band shift due to an increase in 3$d$ orbital occupation by substitution of Co by Ni which shifts the 3$d$ states to higher binding energy.

Finally we comment on the pressure response of Eu$T$Ge$_3$.
Recent temperature dependence of the electrical resistivity studies under pressure reported a successive increase of $T_{\rm N}$ for Eu$T$Ge$_3$ ($T$=Co, Ni, Rh, Ir) \cite{Uchima2014b, Kakihana2017}. The antiferromagnetic ordering in Eu$T$Ge$_3$ stably exists up to 8 GPa and no sign of a Eu valence transition was observed. Our valence band spectra give an explanation to the robust Eu$^{2+}$ magnetic states against pressure. The dominant part of the Eu 4$f$ DOS of Eu$T$Ge$_3$ is localized at $\sim$1 eV below $E_{\rm F}$ which is deeper than other Eu-compounds possessing an intermediate Eu valence state or valence transition \cite{Mimura2004, Ichiki2017}. It hinders a charge transfer from Eu 4$f$ to the conduction band that make up the valence fluctuation or non-magnetic Eu$^{3+}$ states. Moreover, the calculated DOS of Eu$T$Ge$_3$ (see Fig. 3) has a quasi-gap like region just above the $E_{\rm F}$ that can also prevent charge transfer. In divalent antiferromagnetic EuFe$_2$As$_2$, the Eu$^{2+}$ 4$f$ states are localized at 1-2 eV below $E_{\rm F}$ \cite{Adhikary2013}. X-ray absorption spectroscopy under pressure reported a change of antiferromagnetic to ferromagnetic ordering above 8 GPa. Althoug the Eu valence gradually increases, the Eu$^{2+}$ magnetic moments remain up to 20 GPa \cite{Kumar2014}.

\section{Conclusions}
We have performed bulk sensitive HAXPES and successfully revealed the electronic structure of Eu$T$Ge$_3$ ($T$=Co, Ni, Rh and Ir). The Eu 3$d$ core level spectrum revealed that the Eu valence states of all the compounds are almost Eu$^{2+}$ with negligible contribution of Eu$^{3+}$. The estimated Eu valence is close to 2.1. The Ge 2$p$ core level spectrum shows the chemical shift to higher binding energy by changing transition metal from 3$d$ to 4$d$ and to 5$d$ elements indicating the change of chemical bonding between $T$ and Ge. The valence band electronic structure was systematically studied with the support of {\it ab initio} band structure calculations. The experimental valence band spectra shows good accordance with the theoretical simulation. The Eu$^{2+}$ 4$f$ states are localized at $\sim$1 eV below $E_{\rm F}$ in all compounds. All the compounds have a quasi-gap like region just above $E_{\rm F}$. This favours the robust Eu$^{2+}$ magnetic state against transition metal substitution and also gives an explanation to its stability against external pressure.

\acknowledgments
We would like to thank Denis C\'{e}olin and Dominique Prieur for their skilful technical assistance. Y. U. thanks Kojiro Mimura and Arata Tanaka for helpful discussions. P. S. was supported by Ames Laboratory's Laboratory-Directed Research and Development (LDRD) funding. Ames Laboratory is operated for the US Department of Energy by the Iowa State University under Contract No. DE-AC02-07CH11358.


\begin{references}
\bibitem{note1} The Ge 2$p_{3/2}$ peak and its plasmon satellite were fitted by a Gaussian function with FWHM=0.3 eV, and Lorentzian functions with FWHM=0.8 eV and 8 eV, respectively.
\end{references}
\end{document}